\documentclass{elsart}

\usepackage{natbib}
\usepackage{graphicx}
\usepackage{amssymb}

\newcommand\arcdeg{\mbox{$^\circ$}}
\newcommand\arcmin{\mbox{$^\prime$}}
\newcommand\arcsec{\mbox{$^{\prime\prime}$}}
\newcommand\nodata{ ~$\cdots$~ }

\newcommand\aap{A\&A}
\newcommand\apj{ApJ}

\newcommand\apjs{ApJS}
\newcommand\mnras{MNRAS}
\newcommand\pasj{PASJ}
\newcommand\pasp{PASP}

\journal{New Astronomy}

\begin{document}

\begin{frontmatter}

\title{The X-ray Pulse Profile of BG CMi}

\author{Chul-Sung Choi}
\address{International Center for Astrophysics, Korea Astronomy and 
Space Science Institute, 36-1 Hwaam, Yusong, Daejon 305-34, Korea}
\ead{cschoi@kasi.re.kr}

\author{Tadayasu Dotani}
\address{Institute of Space and Astronautical Science, Japan
Aerospace Exploration Agency, 3-1-1 Yoshinodai, Sagamihara,
kanagawa 229-8510, Japan}
\ead{dotani@astro.isas.jaxa.jp}

\author{Yonggi Kim}
\address{University Observatory, Chungbuk National University, 
Cheongju 361-763, Korea}
\ead{ykkim153@chungbuk.ac.kr}

\author{Dongsu Ryu\corauthref{cor}}
\address{Department of Astronomy and Space Science, Chungnam
National University, Daejon 305-764, Korea}
\ead{ryu@canopus.cnu.ac.kr}
\corauth[cor]{Corresponding author.}

\begin{abstract}

We present an analysis of the X-ray data of a magnetic cataclysmic variable,
BG CMi, obtained with ROSAT in March 1992 and with ASCA in April 1996.
We show that four peaks clearly exist in the X-ray pulse profile,
unlike a single peak found in the optical profile.
The fluxes of two major pulses are $\sim 2 - 5$ times larger than those
of two minor pulses.
The fraction of the total pulsed flux increases from 51\% to 85\% with
increasing energy in 0.1 $-$ 2.0~keV, whereas it decreases from 96\% to
22\% in 0.8 $-$ 10~keV.
We discuss the implications of our findings for the origin of the pulse
profile and its energy dependence.

\end{abstract}

\begin{keyword}

cataclysmic variables :  close binaries :  individual stars (BG CMi) :
X-ray binaries

\end{keyword}

\end{frontmatter}

\section{Introduction}

BG CMi is a magnetic cataclysmic variable (mCV), known as an intermediate
polar (IP), with an orbital period of 3.235~hr
\citep[][for review of IPs, see, e.g., Patterson 1994]{mac84}.
The primary star, exhibiting pulsations of fluxes in the X-ray, optical, and
infrared bands, is a rotating magnetic white dwarf.
Its magnetic field strength has been inferred to be
${\rm B} = (2-6) \times 10^6$~G \citep[]{cha90}, based on circular
polarization observations in the optical and infrared bands \citep[]{pen86, wes87}.
However, the fundamental properties of the secondary star are yet poorly known.
The system parameters as well as the distance to the binary system are
not well constrained.

In the optical band, a quasi-sinusoidal pulse profile was clearly
seen at a period of 913.5~s, together with an orbital modulation of
brightness over a period of 3.235~hr \citep[e.g.,][]{mar95, kim05}.
The optical studies claimed that the 913.5~s period is that
of the white dwarf's spin.
According to the pulse period analysis, the primary is spinning up,
but the rate seems to change in time \citep[][and reference therein]{kim05}.
\citet[]{mar95} found that the fraction of the pulsed flux in the optical band
also varies with time and wavelength.

The pulsation and the orbital modulation were also detected in X-ray
at the same periods \citep[]{ter04, par05}.
However, the X-ray pulse profile is non-sinusoidal and hinted possibly
multiple peaks.
\citet[]{mac87} reported that there are two emission components in the
X-ray of a medium energy range, one being pulsed and radiated from the magnetic
poles of the white dwarf and the other being unpulsed and radiated from the
regions where accretion streams impact the magnetosphere.
An iron K$\alpha$ fluorescence line was observed at $\sim 6.4$~keV with
no significant variations over the pulse period \citep[e.g.,][]{nor92, ezu99}.

The study of this paper focuses on the X-ray pulse profile of BG CMi and its
energy dependence from the ROSAT and ASCA archival data.
Based on the analysis results, the geometry of X-ray emitting regions
is discussed.

\section{Observation and Data Reduction}

BG CMi was observed with ROSAT on March 28 to 29, 1992 with
on-source time of $8.4\times10^3$~s.
It was observed twice with ASCA with an interval of 2 days:
April 14 through 15, 1996 with on-source time of $3.7\times10^4$~s
and April 17 through 18, 1996 with on-source time of $3.6\times10^4$~s
\citep[see also][]{ezu99, ter04, par05}.

The ROSAT observation was made with Position Sensitive Proportional
Counter (PSPC-B) mounted on the focal plane of the X-ray telescope.
The PSPC covers the energy range of 0.1 $-$ 2~keV and has a relatively
high spatial resolution of $\sim 25\arcsec$ at 1~keV over 2\arcdeg\ diameter
field of view.
The PSPC is known to have a very high rejection efficiency of particle
background events, 99.9\%, with a typical count rate of $4 \times 10^{-6}$
counts s$^{-1}$ arcmin$^{-2}$ keV$^{-1}$ \citep[]{plu93}.
We acquired the screened data of BG CMi from the HEASARC public archives.

According to the second ROSAT source catalog of pointed observations
with PSPC\footnote{http://wave.xray.mpe.mpg.de/rosat/rra},
other X-ray sources may be present within a circular region of
radius 0.1\arcdeg\ from BG CMi (see Table~1).
For instance, an X-ray source which is relatively bright and has a count
rate of $1.25\times10^{-2}$ counts s$^{-1}$ is located near BG CMi
at the position of R.A. = 112.88126\arcdeg\ and DEC. = 10.06978$\arcdeg$
(J2000) with an angular separation of $7.8\arcmin$.
Caution therefore is needed in an analysis of X-ray data of BG CMi.
Considering this situation, we restrict a source extraction
region to a circle of radius 3\arcmin\ centered at the position
of BG CMi.

ASCA is equipped with four thin-foil X-ray telescopes, focusing X-rays
onto four focal plane detectors, of which two are Solid-state Imaging
Spectrometers (SIS0 and SIS1) and two are Gas Imaging Spectrometers
(GIS2 and GIS3).
Each SIS consists of four CCD chips with an energy resolution of
$\sim\!50 - 160$~eV in the range of 0.4 $-$ 10.0~keV, while GISs have
an energy resolution of $\sim 200 - 600$~eV in 0.8 $-$ 10.0~keV\@.
The telescopes have a $\sim3\arcmin$ half-power diameter of the point
spread function (PSF), and the intrinsic PSF of GISs depends on the
incident X-ray energy and is Gaussian with a FWHM of
$0.5\times \sqrt{5.9\ {\rm keV} / E}$ arcmin (where $E$ is the X-ray
energy in keV).

The ASCA observations were made with both SISs and GISs. SISs
were operated in the 2 CCD faint mode and GISs in the pulse-height (PH) mode.
The time resolution of SISs in the 2 CCD mode is 8~s and that
of GISs in the PH mode is 62.5/500~ms for high/medium telemetry bit rates.
SISs in the 2 CCD mode provide a $11\arcmin \times 23\arcmin$ rectangular
field of view, while GISs give a circular field of view with a
diameter of $50\arcmin$, regardless of their observational mode.

We acquired raw ASCA data from public archives and applied
`strict' data screening criteria to the GIS and SIS data to reduce
possible contaminations from the Earth's bright limb and from the
regions of high particle background.
For example, the SIS data were rejected when the pointing direction of
the telescope is less than 40\arcdeg\ from the Earth's bright limb and
the upper-threshold of the radiation belt monitor was set to 50.
Hot and flickering pixels were removed from the SISs and
a grade-based selection criterion was applied to the data.
Other screening criteria were set to the recommended values
(we checked difference in the GIS data by applying different
screening criteria to the elevation angle, which is defined as the angle
between the source and the Earth's limb, e.g. 5\arcdeg\ and 10$\arcdeg$, but
found no significant difference; we therefore applied the recommended
value to the elevation angle).
After this screening, the source events were extracted from the
3\arcmin\ radius region for both the GISs and SISs data.

\section{Results}

We extracted light curves after converting the X-ray arrival times to
the barycentric times of the solar system.
Therefore, all the dates or times described in this section are
based on the Barycentric Dynamical Time (TDB).
The extracted light curves have a time resolution of 0.1~s for the PSPC
data, 0.5~s for the GISs, and 8.0~s for the SISs.

\subsection{Pulse Period}

To confirm the presence of pulses in the current X-ray data,
we did a period search using an epoch folding method, e.g., `efsearch'
in FTOOLS (or XRONOS).
For this work, we focused on the PSPC and GIS data that have a better time 
resolution than the SIS data.
In this search, we did not correct the Doppler effect due to the
binary motion, because it is negligible in the current data.
Figure~1 shows the result obtained from the GIS data.

The best periods, which were determined by fitting a Gaussian function to
the centroid of $\chi^2$-peak, are $P = 913.6 \pm 0.3$~s for the PSPC data
and $P = 913.5 \pm 0.1$ ~s for the GIS data (where the errors in $\chi^2$
are included).
These periods are consistent with each other within the errors
as well as with those reported in previous X-ray and optical studies.
It is noted from the $\chi^2$-plot that there are higher harmonics of
the period $P$ and a beat between the orbital and pulse periods, marked with
vertical short-dashed lines, suggesting that $P$ is the fundamental period of
the source and is the spin period of the white dwarf.

\subsection{Pulse Profile}

Figure~2 and Figure~3 show the mean pulse profiles (crosses) obtained from
the ROSAT and ASCA data, respectively. The start and stop times of the
data are indicated in each figure.
The ROSAT data were folded at the period of 913.6~s from the epoch
MJD 48709.28492 and the ASCA data were folded at the period of 913.5~s from
the epoch MJD 50187.79824.
The epochs for pulse phase zero were taken at the centers of the highest
pulses (RP1 and AP2), which were determined independently for the ROSAT
and ASCA data through model fits to the profiles (see below).
For this we included the SIS data to increase the data statistics.
(Before we included the SIS data, we inspected visually the pulse
profiles of the four ASCA instruments. We found no significant
difference in, e.g., overall profile, peak positions, and relative pulse height.)

As displayed in Figure~2, the ROSAT data in the 0.1 $-$ 2.0~keV range exhibit two
peaks (or pulses), labeled `RP1' and `RP2', over one cycle of the pulse profile.
The two pulses are not separated by 0.5 in phase and their heights are
not equal. Furthermore, the widths seem to be different too.
The profile can be approximated by a model of two
Gaussians on the top of a constant base, as represented with the solid curve.
On the other hand, the ASCA data in the 0.8 $-$ 10.0~keV range exhibit four pulses
labeled `AP1' $-$ `AP4' in Figure~3, and the profile is reproducible by
a model of four Gaussians on the top of a constant base (see the following subsection for
the model function; we checked whether the profile can be fitted with other
models, e.g., sine plus two Gaussians, but found that none reproduces the
profile successfully).
From the figure, it is noted that the AP1 and AP3 pulses are separated
by $\sim\! 0.5$ in phase and the AP2 and AP4 pulses are separated also by
$\sim\! 0.5$.
However, the overall profiles of AP1/AP3 and AP2/AP4 are
different to each other.
The difference in the ROSAT and ASCA profiles will be described 
in details in the following subsection.

In order to identify which pulse is coincident with the optical pulse maximum, 
we calculated the epoch that is near the start time of the X-ray data,
using the recently presented ephemerides by, e.g., \citet[]{kim05}\footnote{
They presented in the abstract a fourth-order polynomial ephemeris which
is statistically optimal for the optical pulse minimum.
However, the ephemeris should be for the pulse maximum, otherwise it
becomes inconsistent with the observed ephemeris presented in their paper
by a half phase.}, \citet[]{hel97}, and \citet[]{pyc96}.
The vertical short-dashed lines in Figures 2 and 3 represent
the optical pulse maxima calculated from the ephemerides. For the lines we did
not correct the difference between the heliocentric time and the TDB because the
difference is negligibly small ($\lesssim$ 3~s) compared to the pulse period.
Taking into account of the uncertainties in the optical ephemerides,
$\sim 0.02 - 0.03$ in phase, we conclude that RP1 and AP2 correspond to 
the optical pulse maximum.
In what follows, we consider that the pulse phases, originally determined
independently for the ROSAT and ASCA data, are in fact coherently
connected with each other.

\subsection{Energy-Resolved Pulse Profiles}

To understand the difference in the ROSAT and ASCA profiles and
also to see how the profiles vary with energy, we examine the energy-resolved
pulse profiles obtained in the four different energy bands, 0.1 $-$ 0.8~keV,
0.8 $-$ 2.0~keV, 2.0 $-$ 4.0~keV, and 4.0 $-$ 10~keV.
Figure~4 shows the profiles, where the ROSAT and ASCA data were folded
independently using the epochs and periods described in the previous
subsection.
The backgrounds were not subtracted from the pulse profiles of Figure~4,
because it was difficult to acquire the appropriate background data
from the nearby annular region due to the presence of contamination
sources.
However, we note that the backgrounds are generally small compared to the
unpulsed fluxes in Table~2 as estimated below.

The typical fluxes of a blank sky observed with ASCA GIS2 are
$1.3 \times 10^{-3}$ counts s$^{-1}$ in 0.8 $-$ 2.0~keV, $1.1 \times 10^{-3}$
counts s$^{-1}$ in 2.0 $-$ 4.0~keV, and $5.1 \times 10^{-3}$ counts
s$^{-1}$ in 4.0 $-$ 10~keV for the same size as the source extraction region
in the data of cutoff rigidity greater than 4 GeV/c.
The SIS0 data show the fluxes of $3.4 \times 10^{-3}$ counts s$^{-1}$
in 0.8 $-$ 2.0~keV, $1.5 \times 10^{-3}$ counts s$^{-1}$ in 2.0 $-$
4.0~keV, and $1.7 \times 10^{-3}$ counts s$^{-1}$ in 4.0 $-$ 10~keV for
the same size as the source extraction region in the data
of cutoff rigidity greater than 6 GeV/c.
Because BG CMi is located far from the galactic plane, we ignored the contribution 
from the galactic ridge emission \citep[e.g., ][]{kan97}.
These mean fluxes of a blank sky are much smaller than the unpulsed fluxes 
of the ASCA data in 2.0 $-$ 10~keV (Table~2).
However, they are significant in 0.8 $-$ 2.0~keV.
This means that we need to be careful in interpreting the unpulsed flux of the
ASCA data in this lowest energy band.
According to the ROSAT PSPC all-sky survey, the flux of the diffuse X-ray background 
can be roughly estimated to $\sim 10^{-4}$ counts s$^{-1}$ arcmin$^{-2}$ in 
0.1 $-$ 2.0~keV \citep[]{sno95}. It is much smaller than the unpulsed fluxes of
the ROSAT data (Table~2), even if we consider the size of the event extraction region.
We note that the contribution of nearby contamination sources is
also small compared to the unpulsed flux in the ROSAT band.
Therefore, below we argue that the energy-dependent behavior of the pulse
fraction\footnote{Defined as the ratio of the flux in
pulses to the total (pulsed + unpulsed) flux.}
as well as the presence of the unpulsed fluxes are intrinsic to the
source.

To study the pulse parameters quantitatively we chose a model of
two or four Gaussians on the top of a constant base, as follows;
\begin{equation}
f(x) = C + \sum_{i=1}^{\rm 2\ or\ 4}\ N_i\ {\rm
exp}[-(x-P_{i})^2/2W_{i}^2],
\end{equation}
where $f(x)$ is the folded pulse profile, $x$ is the phase, $P$ is
the Gaussian center, $W$ is the width, $N$ is the height, and $C$ is
a constant that represents the unpulsed flux level.
In this model, all the parameters were allowed to be fitted.
The parameters of the best-fit model, represented as solid and
short-dashed curves in Figure~4, are summarized in Table~2.
The mean pulsed flux in each profile is also calculated in the table
using the best-fit parameters.

Although the ROSAT and ASCA data were obtained with an interval of 4
years, their pulse profiles are surprisingly similar.
This means that BG CMi showed very little time variation in 4 years.
One may claim that the pulse profiles in the common energy
band of 0.8 - 2.0 keV look different in the ROSAT and ASCA data
(we checked the ROSAT profile by applying a different folding period,
e.g., 913.5~s, but found no significant difference).
That is, in the ROSAT data of Figure~4b, the two pulses that correspond to
AP1 \& AP4 in the ASCA data of Figure~4c seem to be absent.
We attribute the difference to the different energy
responses of ROSAT PSPC and ASCA SIS/GIS.
PSPC has a larger effective area for softer X-rays in this energy band,
whereas SIS/GIS has a larger effective area for harder X-rays.
In fact, if we check carefully the parameters in Table~2, there is a
hint of AP1 \& AP4 in the ROSAT profile.
For example, the width of RP1 is almost identical to the width of
AP2, whereas the width of RP2 is larger than that of AP3.
This broader pulse may be interpreted as a superposition of two
pulses AP3 and AP4 of the ASCA profile.
In addition, the possible existence of a broad hump above the unpulsed level, which
appears at the phase range $\sim 0.5 - 0.8$ in Figure~4a, may support this
interpretation.
Although we claim the similarity in the pulse profiles of ROSAT and ASCA, it
needs to be confirmed again through observations covering a wide energy range of
0.1 $-$ 10 keV.

By analyzing the energy-resolved pulse profiles together with the fit
parameters in Table~2, we obtained the following results:

\begin{enumerate}

\item  The separation of the AP2 and AP4 pulses, from peak to peak, is sustained to be 
consistent, $\sim 0.5$ in phase, irrespective of energy band.
The AP1 and AP3 pulses are basically separated by $\sim 0.5$, but the
separation tends to increase with increasing X-ray energy.

\item The pulse widths for RP1 and AP2 increase clearly with increasing
energy, from 0.04 to 0.11 in phase.

\item The pulsed fluxes for AP2 and AP3 (major pulses) are $\sim 2 - 5$
times larger than the fluxes for AP1 and AP4 (minor pulses) at all energies.
The flux ratios tend to be lower at higher X-ray energy.
The tendency explains the profile change in the ASCA data, seen in Figure~4c
through Figure~4e.

\item The total pulse fraction
decreases with increasing energy, from 96\% to 22\% in the energy range of
0.8 $-$ 10.0~keV.\footnote{When we calculated the pulse fraction, we
subtracted the backgrounds obtained from the blank sky observation data.}
On the other hand, the pulse fraction in RP1 and RP2 increases from
51\% to 85\% with increasing energy in the energy range of 0.1 $-$ 2.0~keV.
These energy-dependent behavior is mainly associated with the flux
variation of the unpulsed component.

\end{enumerate}

\section{Discussion}

\subsection{Pulse Profile}

As shown in \S3.2, multiple X-ray pulses exist consistently in
0.1 $-$ 10~keV with relatively narrow widths and show no significant
phase shift over the selected energy bands.
These facts suggest that the pulsed radiation originates from
restricted regions of the rotating compact object.
The pulse profile in the ASCA data of 0.8 $-$ 10~keV in Figure~4
contains four pulses, which are particularly well separated in
the lower energy band of 0.8 $-$ 2.0~keV.
The ROSAT data in 0.1 $-$ 2.0~keV also indicate possibly four
pulses.
The four pulse profile requires in general four X-ray emitting
regions on the surface of the white dwarf, if all the pulses are
considered to come from a single object.
In this sense, the postulate that a multipole geometry of magnetic field
exists in the primary star can not be excluded completely \citep[see e.g.,][]{beu07}.

Nevertheless, if the primary star should have a dipole magnetic field geometry, two pulses need to be
generated from a single magnetic pole.
Then we argue that the major pulses (AP2 \& AP3) should be produced from one
magnetic pole, and the minor pulses  (AP1\& AP4) from the opposite
pole, based on the following reasons:
A dipole magnetic field generally produces two pulses
separated by $\sim 0.5$ in phase, unless the dipole moment is severely distorted.
We note that the AP1 and AP3 pulses are separated by
$\sim 0.5$ in phase and the AP2 and AP4 pulses are also separated by
$\sim 0.5$ in phase.
Furthermore, if two pulses are produced by a single
pole, they would have a similar amplitude, because the amplitude is
determined in principle by the geometric effect, such as the occultation
of the emitting region.
However, the existence of deep troughs between the
major pulses and between the minor pulses in Figure~4 implies that the two emission
regions in a pole should be well separated. Such separation favors large magnetic pole
areas (or polar caps), unlike the conventional dipole field geometry of mCVs
that has two small-size polar caps \citep[see e.g.,][]{ros88}.
One interesting point is that the large polar cap hypothesis is also
required to explain the observations of circular polarization in BG CMi
\citep[e.g.,][]{wes87, cha90}.

\subsection{Energy Dependence of Pulse Fraction}

The total pulse fraction decreases clearly from 96\% to 22\%
with increasing X-ray energy in 0.8 $-$ 10.0~keV (\S3.3).
It is also worthwhile to mention that the pulse width
decreases from 0.11 to 0.04 in phase with decreasing energy in RP1 and AP2.
The energy dependence of the pulse fraction may reflect the temperature
distribution in the X-ray emitting regions.
That is, if a relatively hot plasma is formed at a large height above the 
magnetic poles, the plasma may be little occulted by the white dwarf and
will contribute largely to the unpulsed X-ray flux.
On the other hand, if a relatively cool plasma is located just above the magnetic
poles, it may be occulted largely by the white dwarf and will contribute much
to the pulsed X-ray flux.
We therefore conjecture that the magnetically channeled accreting matter 
cools down by radiative cooling processes while it travels toward the magnetic 
pole areas.
This idea may also explain the energy dependence of the pulse width, if we
consider a finite opening angle of the magnetic field.
We leave this issue as a future study because the signal to noise ratios of
the spectral data are not good enough to test the idea.

\subsection{Optical Profile}

It is interesting to note that the optical profile has a
single, quasi-sinusoidal pulse, unlike the X-ray profile we studied.
The total pulse fraction in optical is relatively low, $\lesssim 30$\%,
and varies with wavelength \citep[]{mar95}.
The low pulse fraction as well as the quasi-sinusoidal profile suggest
that the pulsed flux originates from a relatively large area which is
occulted partly and periodically by the rotating white dwarf.

As we showed in \S3.2, the expected optical pulse maximum matches well
with the RP1 and AP2 pulses.
Likewise, the fact that the pulsed flux of the major pulses is
larger by $\sim 5$ times than the flux of the minor pulses indicates that
there exists an appropriate geometrical condition for producing the
quasi-sinusoidal profile.
Therefore, the pulsed flux in the optical band may be suggested to come from
the pole area that was heated up by
the illumination of the X-rays emanating from the channeled accreting matter.
Alternatively, the pulsed flux may originate from the matter
that is cooling and spreading over the white dwarf surface after it
settled on the polar region as discussed by \citet[]{choi00}.

\section{Summary}

Our findings are summarized as follows:

\begin{enumerate}

\item BG CMi has an X-ray profile with four pulses over the period of
913.5~s.

\item The fluxes of the major pulses AP2 and AP3 are $\sim 2 - 5$
times larger than those of the minor pulses AP1 and AP4.

\item The pulse fraction increases from 51\% to 85\% with increasing energy
in 0.1 $-$ 2.0~keV, whereas it decreases from 96\% to 22\% in 0.8 $-$ 10~keV.

\item Unlike the X-ray profile, the optical profile has single, quasi-sinusoidal pulse.
Among the four X-ray pulses, the highest one coincides with the optical
pulse maximum.

\end{enumerate}

\ack

C.S.C. and Y.G.K. thank Dr. I. Andronov for comments 
and discussions on results of this work.
C.S.C. thanks Dr. H. Kim for discussions of magnetic field
configurations in compact stars.
This work was supported in part by the Korea Science \& Engineering Foundation
through the grant of the basic research program R01-2004-000-1005-0.
The work of D.R. was also supported in part by Korea
Foundation for International Cooperation of Science \& Technology
(KICOS) through the Cavendish-KAIST Research Cooperation Center.

\clearpage

\begin{table}
\caption{Possible Cataloged X-ray Sources near BG CMi$^a$}
\begin{tabular}{ccrc}
\hline\hline
Source Name & R.A.$^b$ & DEC.$^b$ &  Count Rate\\
            &          &          &  (counts s$^{-1}$)\\
\hline
2RXP J073138.1+100108 & 112.90876 & 10.01917 & (9.4$\pm$3.7)$\times10^{-4}$\\
2RXP J073126.8+095959 & 112.86167 & 9.99972  & (1.0$\pm$0.4)$\times10^{-3}$\\
2RXP J073148.2+095810 & 112.95084 & 9.96945  & (1.1$\pm$0.4)$\times10^{-3}$\\
BG CMi                & 112.87084 & 9.94028  & (8.7$\pm$0.3)$\times10^{-2}$\\
2RXP J073151.3+095430 & 112.96375 & 9.90833  & (2.0$\pm$0.6)$\times10^{-3}$\\
\hline
\end{tabular}
\\$^a$Within a circular region of radius 0.1\arcdeg\ which
is centered at BG CMi.
\\$^b$Equatorial coordinates are in equinox J2000.
\end{table}

\clearpage

\begin{table}
\begin{small}
\caption{Best-Fit Pulse Parameters and Calculated Pulsed fluxes}
\begin{tabular}{lcccc}
\hline\hline
\multicolumn{1}{l}{Pulse or Peak} &
\multicolumn{4}{c}{Energy Band} \\ \cline{2-5}
        & 0.1 $-$ 0.8 keV & 0.8 $-$ 2.0 keV & 2.0 - 4.0 keV & 4.0 $-$ 10 keV\\
\hline
RP1 & & & \\
$P$ (phase)            &  0.995$^{+0.005}_{-0.006}$ & 
1.00$^{+0.02}_{-0.01}$
                        & \nodata & \nodata\\
$W$ (phase)            &  0.036$^{+0.005}_{-0.006}$ & 
0.065$^{+0.010}_{-0.010}$
                        & \nodata & \nodata\\
$Flux^{*1}$ ($10^{-3}$ counts s$^{-1}$) &  21.5$^{+8.3}_{-6.4}$ & 
18.3$^{+7.1}_{-5.5}$
                                         & \nodata & \nodata\\
\hline
RP2 & & & \\
$P$ (phase)            &  1.27$^{+0.01}_{-0.01}$    & 
1.28$^{+0.02}_{-0.02}$
                        & \nodata & \nodata\\
$W$ (phase)               &  0.06$^{+0.01}_{-0.01}$ & 
0.10$^{+0.02}_{-0.02}$
                        & \nodata & \nodata\\
$Flux^{*1}$ ($10^{-3}$ counts s$^{-1}$) &  15.0$^{+7.1}_{-5.3}$ & 
17.7$^{+9.1}_{-6.3}$
                                         & \nodata & \nodata\\
\hline
$C^{*2}$ ($10^{-3}$ counts s$^{-1}$)      &  34.9  & 6.4 & \nodata & 
\nodata\\
\hline
AP1 & & & \\
$P$ (phase)            & \nodata & 0.77$^{+0.01}_{-0.01}$ & 
0.76$^{+0.01}_{-0.01}$
                        & 0.76$^{+0.02}_{-0.02}$\\
$W$ (phase)            & \nodata & 0.067$^{+0.013}_{-0.010}$ & 
0.055$^{+0.010}_{-0.007}$
                                  & 0.06$^{+0.02}_{-0.02}$\\
$Flux^{*1}$ ($10^{-3}$ counts s$^{-1}$) & \nodata & 
1.6$^{+0.6}_{-0.4}$
                                         & 3.6$^{+0.9}_{-0.8}$ & 
2.4$^{+1.7}_{-1.0}$\\
\hline
AP2 & & & \\
$P$ (phase)            & \nodata & 1.016$^{+0.004}_{-0.005}$ & 
1.009$^{+0.007}_{-0.008}$
                        & 1.01$^{+0.02}_{-0.02}$\\
$W$ (phase)            & \nodata & 0.068$^{+0.006}_{-0.006}$ & 
0.088$^{+0.010}_{-0.009}$
                        & 0.11$^{+0.03}_{-0.03}$\\
$Flux^{*1}$ ($10^{-3}$ counts s$^{-1}$) & \nodata & 
5.2$^{+0.8}_{-0.8}$
                                      & 10.5$^{+1.9}_{-1.7}$ & 
6.1$^{+3.3}_{-2.0}$\\
\hline
AP3 & & & \\
$P$ (phase)             & \nodata & 1.282$^{+0.003}_{-0.013}$ & 
1.32$^{+0.03}_{-0.04}$
                         & 1.36$^{+0.02}_{-0.04}$ \\
$W$ (phase)             & \nodata & 0.075$^{+0.012}_{-0.017}$ & 
0.09$^{+0.02}_{-0.02}$
                         & 0.09$^{+0.02}_{-0.03}$\\
$Flux^{*1}$ ($10^{-3}$ counts s$^{-1}$) & \nodata & 
4.8$^{+1.0}_{-2.5}$
                                         & 9.8$^{+3.3}_{-3.6}$ & 
5.3$^{+2.4}_{-2.6}$\\
\hline
AP4 & & & \\
$P$ (phase)             & \nodata & 1.50$^{+0.04}_{-0.16}$ & 
1.49$^{+0.02}_{-0.03}$
                         & $1.50^{*3}$\\
$W$ (phase)             & \nodata & 0.10$^{+0.10}_{-0.03}$  & 
0.06$^{+0.02}_{-0.03}$
                         & $0.036^{*3}$\\
$Flux^{*1}$ ($10^{-3}$ counts s$^{-1}$) & \nodata & 
1.7$^{+2.3}_{-0.7}$
                                         & 2.3$^{+2.9}_{-0.4}$ & 0.7\\

\hline
$C^{*2}$ ($10^{-3}$ counts s$^{-1}$)      & \nodata & 2.8 & 20.5 & 52.4\\

\hline
\end{tabular}
\\Note - Models of two or four Gaussians on the top of a constant base were fitted to the pulse
profiles. 
The full-width at half-maximum can be obtained by multiplying a constant,
i.e., $W_{\rm FWHM} = 2.354 \times W$.
All the attached errors are at 90\% confidence level.
\\$^{*1}$The
mean pulsed flux for each pulse was calculated using the best-fit parameters.
\\$^{*2}$A constant that represents the unpulsed flux level.
\\$^{*3}$Fixed in the fit.
\end{small}
\end{table}

\clearpage

\begin{figure}
\includegraphics[angle=90,scale=0.555]{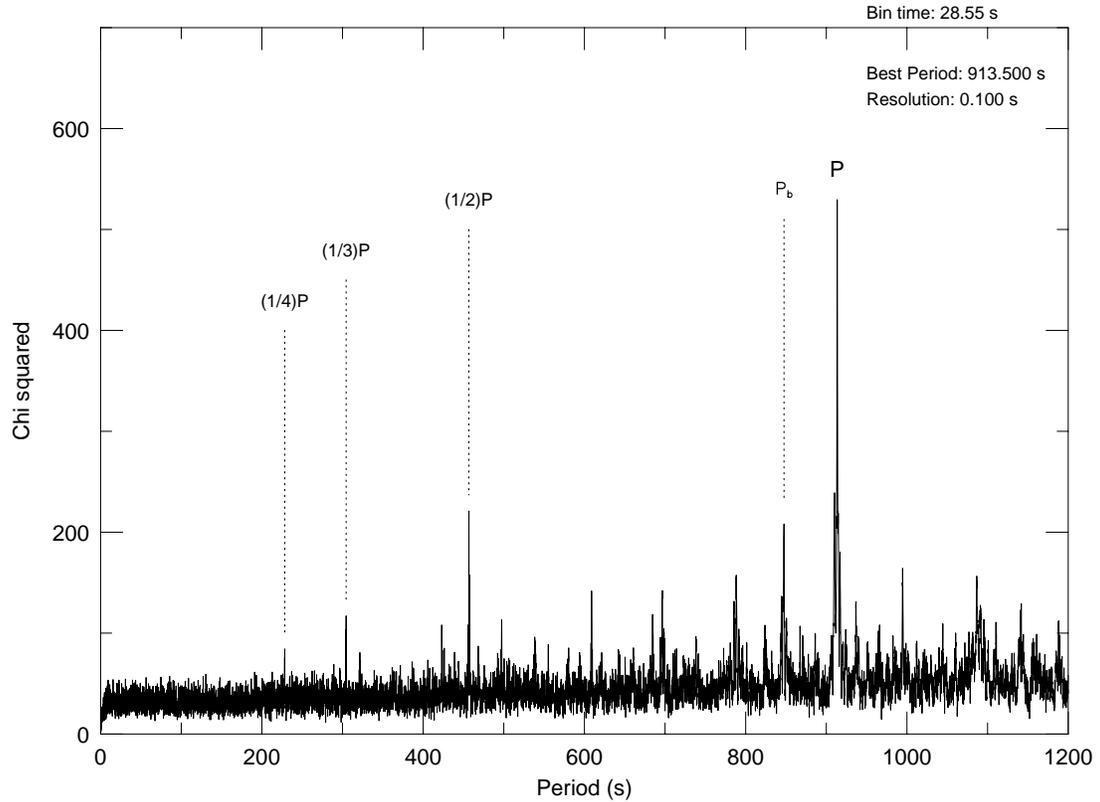}
\caption{
$\chi^2$-plot as a function of trial period for the GIS data in the entire
energy range.
The harmonics of the period P, labeled with $1 \over 2$P (456.7~s), 
$1 \over 3$P (304.5~s), and $1 \over 4$P (228.4~s), are clearly seen in
the figure. 
The peak labeled with P$_{\rm b}$
($\equiv$ ${\rm P_{orb}P}\over{\rm P_{orb} + P}$ = 847.5~s) is the beat between
orbital (3.235~hr) and pulse (913.5~s) periods.
}
\end{figure}

\clearpage

\begin{figure}
\includegraphics[angle=0,scale=0.6]{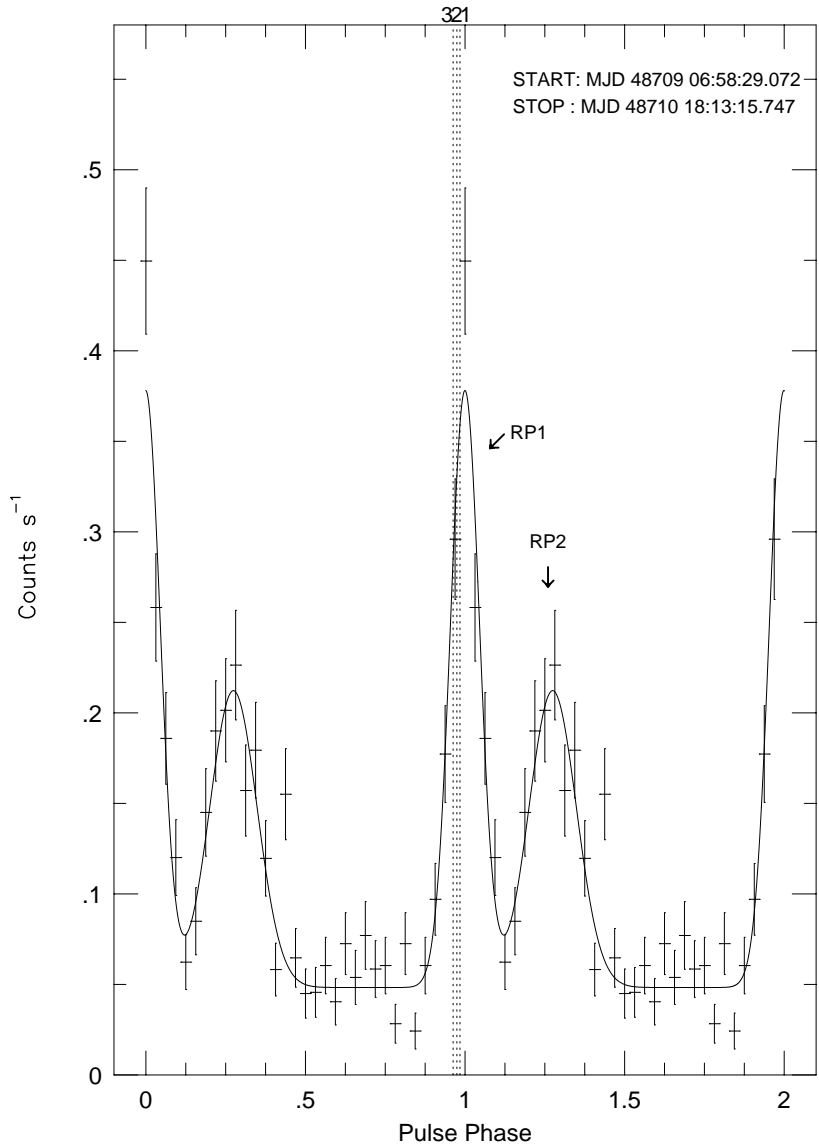}
\caption{
Pulse profile of the ROSAT data folded at period 913.6~s from
the epoch MJD 48709.28492. The pulse phase has been repeated over two cycles.
A model of two Gaussians on the top of a constant base (solid curve) was fitted to the
profile (crosses) from the minimum to the minimum. 
The vertical short-dashed lines represent the optical pulse maximum expected
from the recent ephemerides: the number 1 at the 
top of the figure is from Kim et al. (2005), 2 is from Pych et al. (1996),
and 3 is from Hellier (1997).
}
\end{figure}

\clearpage

\begin{figure}
\includegraphics[angle=0,scale=0.6]{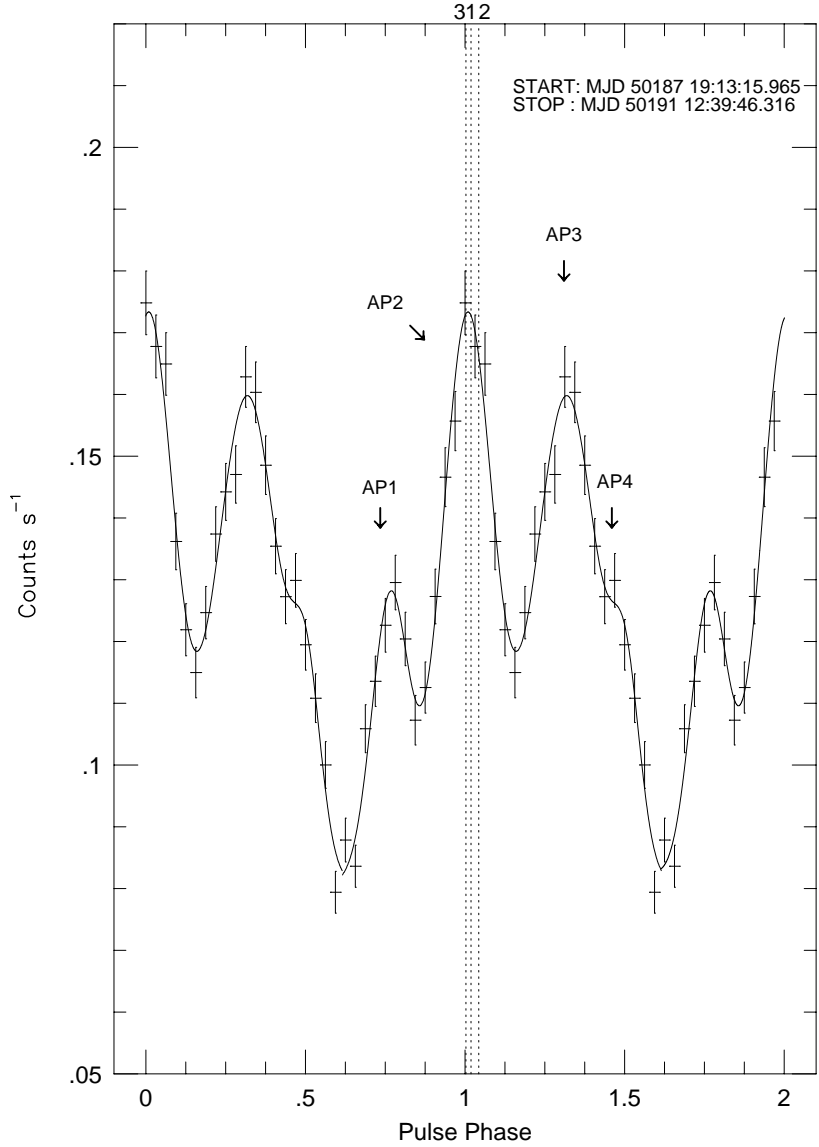}
\caption{
Pulse profile of the ASCA data folded at period 913.5~s from
the epoch MJD 50187.79824.
The count rate is an average value detected by GISs and SISs.
A model of four Gaussians on the top of a constant base (solid curve) was fitted to
the profile (crosses) from the minimum to the minimum. 
The vertical short-dashed lines are the same as in Figure~2.
}
\end{figure}

\clearpage

\begin{figure}
\includegraphics[angle=0,scale=0.7]{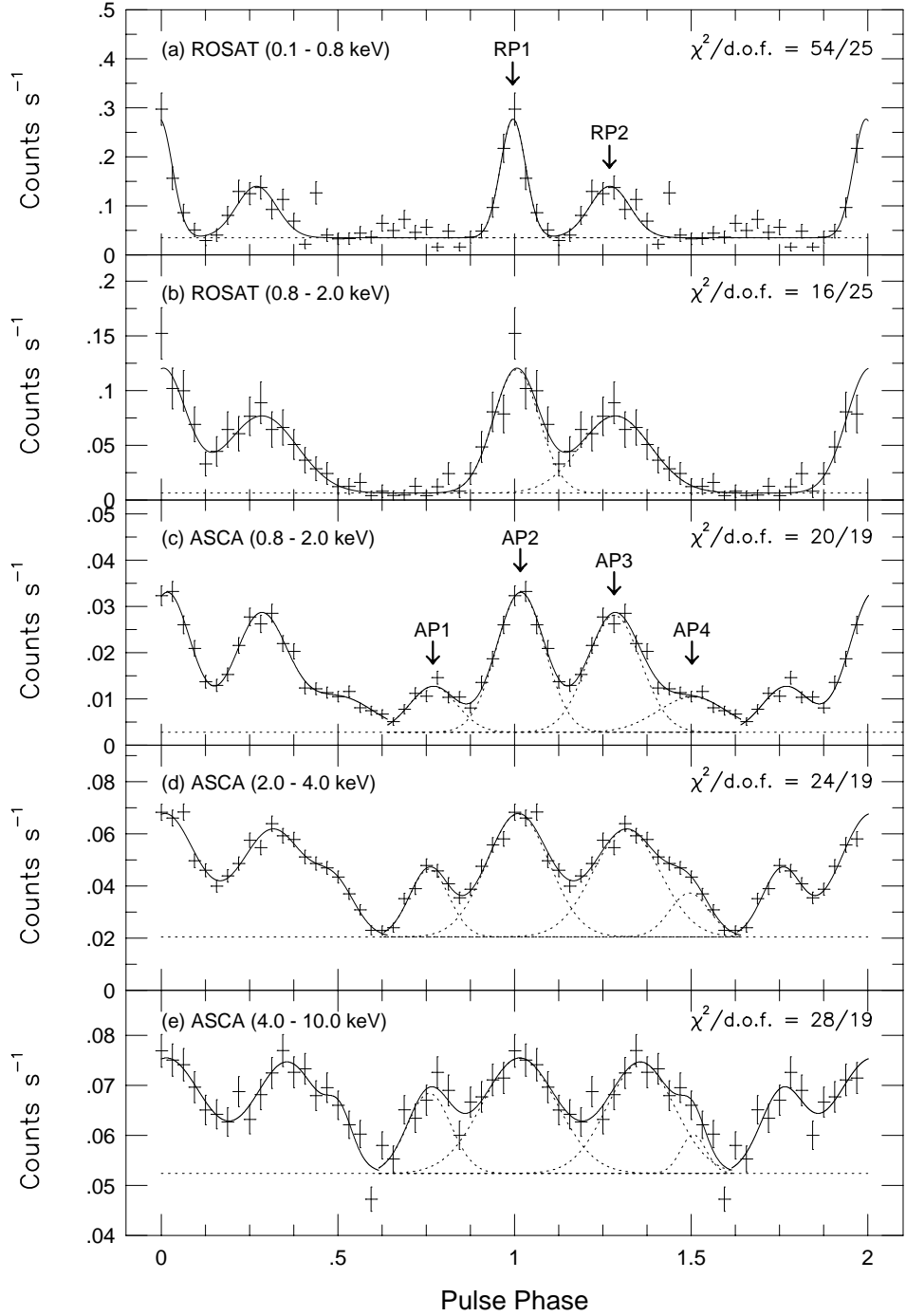}
\caption{
Energy-resolved pulse profiles of BG CMi.
The folding epoch and period for the ROSAT and ASCA data are the same as
in Figures~2 and 3, respectively.
The short-dashed curves and horizontal lines represent each Gaussian
component and the unpulsed flux level, respectively.
Goodness of the model-fit is inserted in each figure.
The count rate in (c) through (e) is an average value
detected by GISs and SISs.
}
\end{figure}

\end{document}